\documentclass[prepreint,twocolumn]{aastex62}
\hypersetup{linkcolor={blue},citecolor={blue}}
\received{}
\revised{}
\accepted{\today}

\shorttitle{Central-engine powered SN ejecta in 3D}
\shortauthors{Suzuki \& Maeda}

\begin{document}
\title{Three-dimensional hydrodynamic simulations of supernova ejecta with a central energy source}

\correspondingauthor{Akihiro Suzuki}
\email{akihiro.suzuki@nao.ac.jp}

\author[0000-0002-7043-6112]{Akihiro Suzuki}
\affil{Division of Science, National Astronomical Observatory of Japan, 2-21-1 Osawa, Mitaka, Tokyo 181-8588, Japan}

\author{Keiichi Maeda}
\affiliation{Department of Astronomy, Kyoto University, Kitashirakawa-Oiwake-cho, Sakyo-ku, Kyoto, 606-8502, Japan}



\begin{abstract}
We present the results of three-dimensional special relativistic hydrodynamic simulations of supernova ejecta with a powerful central energy source. 
We assume spherical supernova ejecta freely expanding with the initial kinetic energy of $10^{51}$ erg. 
We performed two simulations with different total injected energies of $10^{51}$ and $10^{52}$ erg to see how the total injected energy affects the subsequent evolution of the supernova ejecta. 
When the injected energy well exceeds the initial kinetic energy of the supernova ejecta, the hot bubble produced by the additional energy injection overwhelms and penetrates the whole supernova ejecta, resulting in clumpy density structure. 
For the smaller injected energy, on the other hand, the energy deposition stops before the hot bubble breakout occurs, leaving the outer envelope well-stratified. 
This qualitative difference may indicate that central engine powered supernovae could be observed as two different populations, such as supernovae with and without broad-line spectral features, depending on the amount of the total injected energy with respect to the initial kinetic energy. 
\end{abstract}



\section{INTRODUCTION\label{intro}}
There is growing evidence that several classes of unusual core-collapse supernovae (CCSNe) harbor a powerful ``engine'' at the center. 
Broad-lined Ic SNe (SNe Ic-BL) are characterized by broad absorption features in their spectra, indicating large explosion energies. 
It is widely known that some SNe Ic-BL are associated with long-duration gamma-ray bursts (GRBs) and thus require an ultra-relativistic jet penetrating the star \citep{2006ARA&A..44..507W,2012grbu.book..169H,2017AdAst2017E...5C}. 
These energetic CCSNe are also recognized as bright radio emitters, which strongly indicates the presence of a fast ejecta component and a (mildly) relativistic blast wave giving rise to radio synchrotron emission \citep[e.g.,][]{1998Natur.395..663K,2010Natur.463..513S}. 
In addition, one of the most plausible scenarios for the emerging new class of hydrogen-poor superluminous SNe (also known as type I superluminous SNe, hereafter SLSNe-I; \citealt{2007ApJ...668L..99Q,2009ApJ...690.1358B,2010ApJ...724L..16P,2011Natur.474..487Q,2011ApJ...743..114C}) is the so-called central engine scenario, in which the central compact object, a highly rotating magnetized neutron star \citep{2010ApJ...717..245K,2010ApJ...719L.204W} or a black hole accretion disk \citep{2013ApJ...772...30D}, injects an additional energy into the surrounding supernova ejecta to give rise to bright thermal emission \citep[see, e.g.,][for review]{2011Natur.474..487Q,2012Sci...337..927G,2018SSRv..214...59M}. 
Despite a lot of investigations on the effects of the putative central engine left in SN ejecta, there are still many problems remained unanswered, such as, how it is produced, how exactly it deposits an additional energy into the surrounding SN ejecta, and what kind of conditions should be met for their progenitor systems. 

Recently, SLSNe and their origin have been paid great attention. 
Despite their small occurrence rate \citep[e.g.][]{2013MNRAS.431..912Q,2015MNRAS.448.1206M,2017MNRAS.464.3568P}, their extreme brightness in optical and UV bands makes it feasible to find them even in distant galaxies \citep{2012Natur.491..228C,2012MNRAS.422.2675T,2013MNRAS.435.2483T,2014ApJ...796...87I,2018A&A...609A..83I}. 
The recent discoveries of spectroscopically confirmed SLSNe at $z\sim 2$ by Dark Energy Survey Supernova Program \citep{2017MNRAS.470.4241P,2018ApJ...854...37S} and Subaru/Hyper Sprime-Cam (HSC) \citep{2019ApJS..241...16M,2019ApJS..241...17C} indeed demonstrated their potential as a probe of star-forming activity in high-z universe. 

However, the power source(s) of the bright emission of SLSNe-I is still unclear. 
In addition to the central engine scenario introduced above, there are two competing scenarios; the interaction between the circumstellar medium and the SN ejecta (CSM scenario; \citealt{2011ApJ...729L...6C,2012ApJ...757..178G,2013MNRAS.428.1020M}), and pair-instability SNe (PISN scenario; \citealt{1967PhRvL..18..379B,1967ApJ...148..803R,2002ApJ...567..532H,2007Natur.450..390W,2009Natur.462..624G}). 
These three scenarios have both advantages and disadvantages in explaining the whole populations of SLSNe-I. 
Therefore, whether the observed population of SLSNe-I can further be divided into two or more subclasses and whether SLSNe-I share some common properties with other types, are important questions ever since SLSNe were recognized as a distinguishing population. 
\cite{2012Sci...337..927G} proposed the presence of the so-called SLSNe-R subclass; SLSNe-I whose late-time light curves can be accounted for by radioactive $^{56}$Ni.
The required amount of $^{56}$Ni is huge, $1$--$10\ M_\odot$, and therefore is unlikely synthesized in normal CCSNe, but may be in line with very massive bare carbon-oxygen cores invoked in the PISN scenario. 
However, the PISN scenario is also known to have difficulty in explaining some rapidly evolving SLSNe-I because theoretical PISN models predict much longer photon diffusion times \citep{2013Natur.502..346N}. 
Furthermore, they do not succeed in reproducing blue spectra of SLSNe-I \citep[e.g.,][]{2012MNRAS.426L..76D,2016MNRAS.455.3207J}. 
In addition, there are some observational studies suggesting the presence of fast- and slowly-evolving populations of SLSNe-I (e.g., \citealt{2013ApJ...770..128I,2017MNRAS.468.4642I}, see also \citealt{2015MNRAS.452.3869N,2017ApJ...850...55N,2017MNRAS.469.1246K}), which might indicate two distinct channels or progenitors for SLSNe-I. 

Many authors have provided model light curves of SLSNe-I \citep{2013ApJ...770..128I,2013ApJ...773...76C,2015MNRAS.452.3869N,2015ApJ...799..107W,2015ApJ...807..147W,2016ApJ...821...22W,2017ApJ...837..128W,2017ApJ...850...55N}, mostly based on the standard treatment of thermal photon diffusion in spherically expanding ejecta (Arnett's solution; see, \citealt{1980ApJ...237..541A,1982ApJ...253..785A,1996snih.book.....A}). 
However, photometric data alone appear to be difficult to distinguish one scenario from the others. 
Recently, thanks to the currently ongoing transient survey programs, such as the Zwicky Transient Facility \citep[ZTF;][]{2014htu..conf...27B}, Pan-STARRS \citep{2002SPIE.4836..154K}, and HSC Subaru Strategic Program, there are plenty of observational data of SLSNe-I accumulated \citep[e.g.,][]{2018ApJ...860..100D,2018ApJ...852...81L}. 
With these rich photometric and spectroscopic data sets, some authors have performed analyses with more sophisticated statistical approaches \citep[e.g.,][]{2017ApJ...845...85L,2017ApJ...850...55N}. 
Most recently, \cite{2018ApJ...854..175I} developed a statistical method to distinguish and classify SLSNe-I by using photometric datasets combined with Gaussian process regression. 
They claim that there are two distinct populations characterized by fast and slow evolutions. 
\cite{2018ApJ...855....2Q} have also developed their own method for distinguishing SLSNe-I from other types by systematically analyzing light curves and spectra of SNe from the Palomar Transient Factory (PTF) archival data. 
They suggest that their SLSN-I samples can be divided into two groups, PTF12dam-like and SN2011ke-like objects, according to their spectral similarities. 
The latter objects show smoother spectra around the optical maximum than the former objects. 
\cite{2019ApJ...871..102N} performed a principal component analysis of the nebular spectra of SLSN-I samples and found no firm evidence of sub-populations. 

On the other hand, SLSNe-I are known to share some common properties with SNe Ic-BL and GRB-SNe in terms of spectroscopic features \citep{2010ApJ...724L..16P,2016ApJ...828L..18N,2017ApJ...835...13J,2017ApJ...845...85L} and host galaxies \citep{2011ApJ...727...15N,2014ApJ...787..138L,2015MNRAS.449..917L,2015MNRAS.451L..65T,2016MNRAS.458...84A,2016ApJ...830...13P,2017MNRAS.470.3566C,2018MNRAS.473.1258S,2018ApJ...857...72H}. 
Moreover, the luminous SN component superposed on the afterglow emission of the ultra-long GRB 111209A, dubbed SN 2011kl \citep{2015Natur.523..189G,2019A&A...624A.143K}, was brighter than any other SNe associated with GRBs and was remarkably similar to SLSNe-I. 
These similarities may further support the idea that at least some SLSNe-I are actually powered by a central energy source and motivate some authors to put forward unified scenarios for energetic and bright CCSNe \citep[e.g.,][]{2015MNRAS.454.3311M,2016ApJ...818...94K}. 
If both SLSNe-I and SNe Ic-BL are powered by a central engine, the engine should play multiple roles. 
In other words, it sometimes powers bright optical emission lasting up to hundreds of days, while it produces energetic ejecta with large kinetic energies or even launches an ultra-relativistic jet. 
It is still unclear whether a single scenario covers these distinct time scales. 
On one hand, the two different energy deposition mechanisms are realized in a single object with different settings. 
For example, \cite{2018MNRAS.475.2659M} suggest that the inclination of the dipole magnetic field with respect to the rotational axis of a rotating magnetized proto-neutron star may lead to both GRB-like jet and quasi-spherical wind. 
On the other hand, this diversity may be explained by physically distinct central engines, neutron stars for a population and black holes for the other. 
Currently, what the central engine is and how exactly it deposits energy into the supernova ejecta are still unclear.

How the deposited energy affects the dynamical evolution of supernova ejecta is of critical importance in observations of CCSNe. 
Properties of supernova ejecta with a central energy source have been investigated mainly in the context of Galactic pulsar wind nebula embedded in supernova remnants (e.g., \citealt{1984ApJ...280..797C,1992ApJ...395..540C,1998ApJ...499..282J}, see \citealt{2006ARA&A..44...17G}, for a review). 
Recently, this issue attracts further attention in the context of the central engine model for SLSNe-I. 
However, it is only recently that multi-dimensional numerical simulations are performed in this context \citep{2016ApJ...832...73C,2017MNRAS.466.2633S,2017ApJ...845..139B}. 
In our previous study (\citealt{2017MNRAS.466.2633S}; hereafter, \hyperlink{sm17}{SM17}), we performed a long-term hydrodynamic simulation of SN ejecta with a relativistic wind injected from the center in 2D cylindrical geometry. 
The simulation revealed that hydrodynamic instabilities caused by the ejecta-wind interaction play a vital role in determining the resultant ejecta structure. 
However, the assumed axisymmetry resulted in an artificial bipolar structure in the density distribution of the ejecta in the free expansion stage. 
Therefore, in this work, we perform simulations with similar numerical settings in 3D cartesian coordinates. 
We find that the dynamical evolution of the ejecta in the 3D simulation is qualitatively similar to the previous 2D simulation and find that the analytical consideration in \hyperlink{sm17}{SM17} can also apply to the ejecta in 3D. 
Furthermore, we also consider the case in which the total amount of the injected energy is comparable to the original kinetic energy of the ejecta. 

This paper is organized as follows. 
In Section \ref{sec:numerical_setups}, we describe numerical setups. 
We perform a couple of numerical simulations with different amounts of the injected energy. 
Results of the numerical simulations are presented and compared in Section \ref{sec:results}. 
We discuss some implications of the numerical results on properties of SN ejecta with a central energy source in Section \ref{sec:discussions}. 
Finally, Section \ref{sec:conclusions} concludes this paper. 

\section{Numerical Setups}\label{sec:numerical_setups}
The numerical setups of the 3D simulation are similar to \hyperlink{sm17}{SM17} for the purpose of comparison. 
In the following, we briefly describe the assumptions and conditions of our 3D simulations. 

\subsection{Supernova ejecta}
We assume spherical SN ejecta with the mass $M_\mathrm{ej}=10M_\odot$ and the kinetic energy $E_\mathrm{sn}=10^{51}$ erg. 
The radial velocity $v$ of a layer is given by its radius $r$ divided by the elapsed time $t$, $v=r/t$ for $v<v_\mathrm{max}$. 
The evolution of the ejecta starts at $t=t_0$. 
The radial density profile $\rho(r,t)$ at $t=t_0$ is described by a broken power-law function,
\begin{equation}
\rho(r,t_0)=
\left\{
\begin{array}{ccl}
\rho_0\left(\frac{v}{v_\mathrm{br}}\right)^{-\delta}&\mathrm{for}&v\leq v_\mathrm{br},\\
\rho_0\left(\frac{v}{v_\mathrm{br}}\right)^{-m}&\mathrm{for}&v_\mathrm{br}<v\leq v_\mathrm{max}, 
\end{array}
\right.
\end{equation}
\citep{1989ApJ...341..867C,1999ApJ...510..379M}. 
The velocity at the break $v_\mathrm{br}$ separates the inner and outer parts of the ejecta. 
The non-dimensional parameters $\delta$ and $m$ specify the inner and outer density slopes. 
The break velocity is expressed in terms of the ejecta mass and the kinetic energy as follows;
\begin{equation}
v_\mathrm{br}=\sqrt{\frac{2f_5E_\mathrm{sn}}{f_3M_\mathrm{ej}}},
\label{eq:vmax}
\end{equation}
with 
\begin{equation}
f_l=\frac{(m-l)(l-\delta)}{m-\delta-(l-\delta)(v_\mathrm{br}/v_\mathrm{max})^{m-l}}.
\end{equation}
The density normalization constant $\rho_0$ is given by
\begin{equation}
\rho_0=\frac{f_3M_\mathrm{ej}}{4\pi v_\mathrm{br}^3t_0^3}.
\end{equation}
For centrally concentrated ejecta with a steep outer slope (i.e., a small $v_\mathrm{br}/v_\mathrm{max}$ and a large $m$) as assumed in our simulations, the numerical factor $f_l$ can be approximated as,
\begin{equation}
f_l=\frac{(m-l)(l-\delta)}{m-\delta}. 
\end{equation}
Accordingly, Equation (\ref{eq:vmax}) yields,
\begin{equation}
v_\mathrm{br}=\sqrt{\frac{2(m-5)(5-\delta)E_\mathrm{sn}}{(m-3)(3-\delta)M_\mathrm{ej}}}.
\end{equation}
We assume the same inner and outer slopes as \hyperlink{sm17}{SM17}, $\delta=1$ and $m=10$. 
The break velocity yields $v_\mathrm{br}=3.8\times 10^8$ cm s$^{-1}$ with this parameter set. 
We further assume that the maximum velocity is 10 times larger than the break velocity, $v_\mathrm{max}=3.8\times 10^9$ cm s$^{-1}$. 

\subsection{Energy injection}
The energy injection at the center of the ejecta is realized as follows. 
We specify a small spherical region at the center of the numerical domain, in which we deposit an additional energy. 
In the energy injection region, thermal gas is injected at a constant energy injection rate $L$ and then the injection is terminated at $t=t_\mathrm{term}$, when the total injected energy reaches $E_\mathrm{inj}=L(t_\mathrm{term}-t_0)$. 
The corresponding mass injection rate is denoted by $\dot{M}c^2=L/\Gamma_\mathrm{cr}$, where $\Gamma_\mathrm{cr}$ is a parameter governing the baryon richness of the injected gas and set to $\Gamma_\mathrm{cr}=20$. 
This value indicates that the rest mass energy of the injected gas is smaller than the total injected energy by a factor of $20$. 
This baryon richness is so small that the injected gas is relativistic. 
In the previous 2D simulation (\hyperlink{sm17}{SM17}), we adopted a fixed energy injection region. 
In 3D simulations, however, it is numerically expensive to keep resolving the fixed energy injection region as its radius progressively becomes small compared with the physical scale of the ejecta. 
Therefore, we adopt an expanding energy injection region. 
The energy injection radius linearly increases with time according to the free expansion,
\begin{equation}
R_\mathrm{in}(t)=R_\mathrm{in,0}\left(\frac{t}{t_\mathrm{0}}\right),
\end{equation}
where $R_\mathrm{in,0}=10^{11}$ cm is the initial values of the energy injection radius. 
The energy injection rate is set to $L=10^{46}$ erg s$^{-1}$, following \hyperlink{sm17}{SM17}. 

We carry out a couple of simulations with different amounts of the injected energy. 
In model E52, we continue the energy injection until the total amount of the injection energy reaches $E_\mathrm{inj}=10^{52}$ erg at $t=t_0+10^6$ s. 
This model is the 3D counterpart of our previous 2D cylindrical simulation. 
We perform another simulation with the total injected energy of $E_\mathrm{inj}=10^{51}$ erg at $t=t_0+10^5$ s (referred to as model E51), while the energy injection rate is same as model E52. 
In this model, the injected energy is comparable to the initial kinetic energy of the supernova ejecta. 
In the following, we compare these two models to see how the dynamical evolution of the supernova ejecta depends on the amount of the injected energy. 

It is convenient to normalize the elapsed time by the characteristic time scale of the energy injection, $t_\mathrm{c}=E_\mathrm{sn}/L=10^5$ s. 
We begin our simulations at $t=t_\mathrm{0}=0.02t_\mathrm{c}$ and evolve the system up to $t=20t_\mathrm{c}$. 

\subsection{Adaptive mesh refinement and coarsening}
Our special relativistic hydrodynamics code is equipped with an adaptive mesh refinement technique (AMR; \citealt{1989JCoPh..82...64B}), which automatically covers regions requiring finer resolutions by smaller grids. 
In particular, shock waves driven by the outermost layer of the ejecta and the hot bubble created by the energy injection are captured by numerical cells with relatively fine resolutions. 
However, tracking the propagation of the shock waves with a fixed resolution requires an impractically large number of numerical cells to cover their structures, especially, in 3D simulations. 
Therefore, in order to ensure realistic computational costs, we eventually relax the numerical resolution by gradually reducing the maximum level of the AMR grid. 
As a result, the minimum resolved length increases with time. 
Nevertheless, it is always kept small compared to the physical scale of the entire ejecta. 
The base grid, i.e., the AMR level $l=0$, covers the numerical domain of $-4.8\times 10^{16}\mathrm{cm}\leq x,y,z\leq 4.8\times 10^{16}\mathrm{cm}$ by $64^3$ numerical cells. 
The maximum AMR level is initially set to $l=16$, corresponding to the numerical domain effectively covered by $\sim(4.2\times 10^6)^3$ cells. 
Despite the slightly different treatments of the energy injection and the coarsening of the numerical grid, we obtain qualitatively similar results to the 2D counterpart as we shall see below.

\section{Results}\label{sec:results}
In the following, we present the results of the numerical simulations. 

\subsection{Model E52}\label{sec:E52_model}
\subsubsection{Hot bubble expansion and breakout}\label{sec:hot_bubble_expansion_and_breakout}
\begin{figure*}
\begin{center}
\includegraphics[scale=1.2,bb= 0 0 446 502]{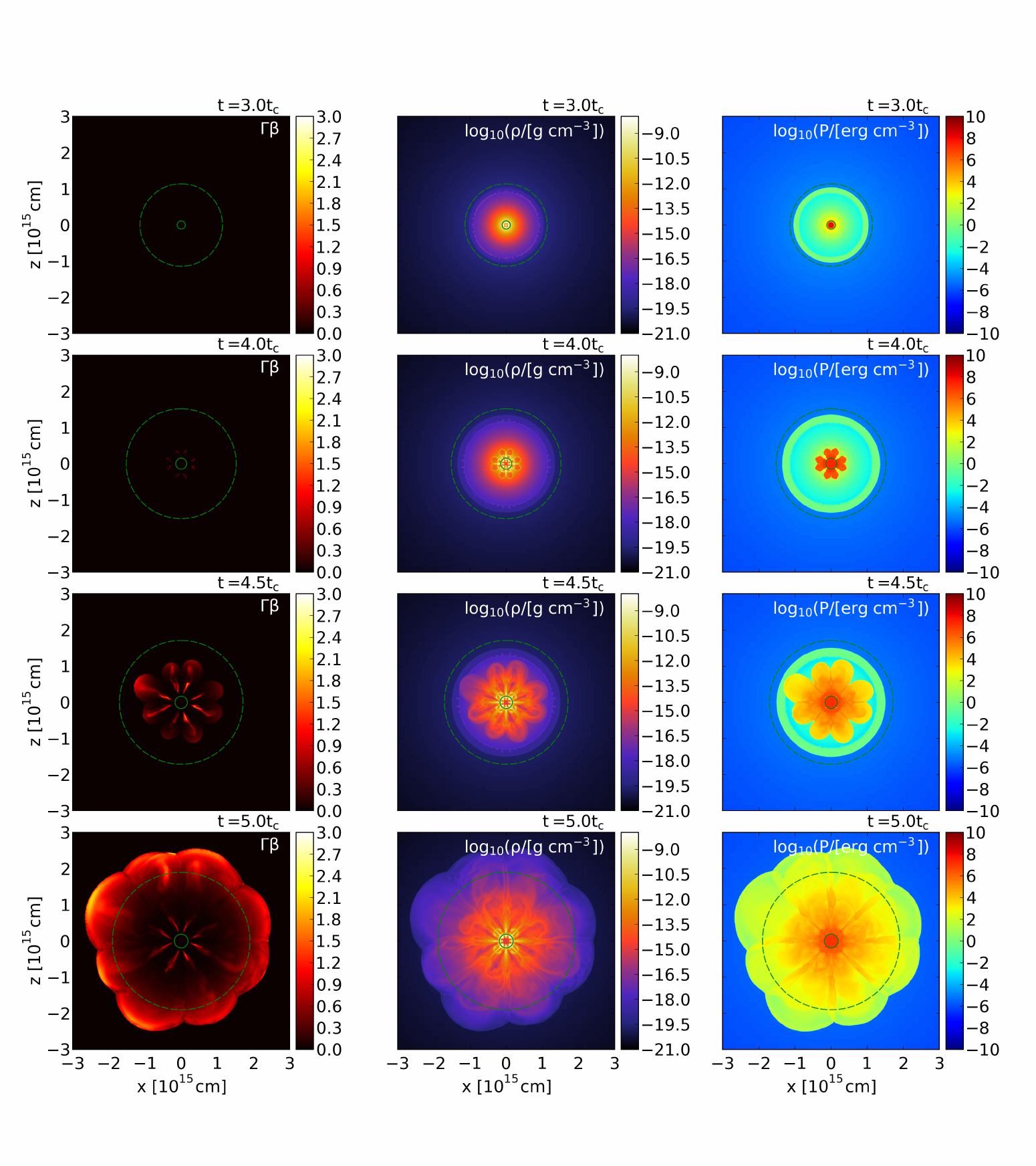}
\caption{Spatial distributions of physical variables at $t/t_\mathrm{c}=3.0$, $4.0$, $4.5$, and $5.0$ (from top to bottom) for model E52. 
The left, center, and right columns represent the spatial distributions of the 4-velocity $\Gamma\beta$, density, and pressure.
The hot bubble initially confined in the inner ejecta destroys the surrounding shell and then breaks out from the outermost layer.
The inner and outer green circles (solid and dashed lines) correspond to the radius of the interface separating the inner and outer ejecta, $r=v_\mathrm{br}t$, and the maximum ejecta radius without the interaction with the ambient matter, $r=v_\mathrm{max}t$.} 
\label{fig:breakout}
\end{center}
\end{figure*}

\begin{figure}
\begin{center}
\includegraphics[scale=0.55,bb= 0 0 432 504]{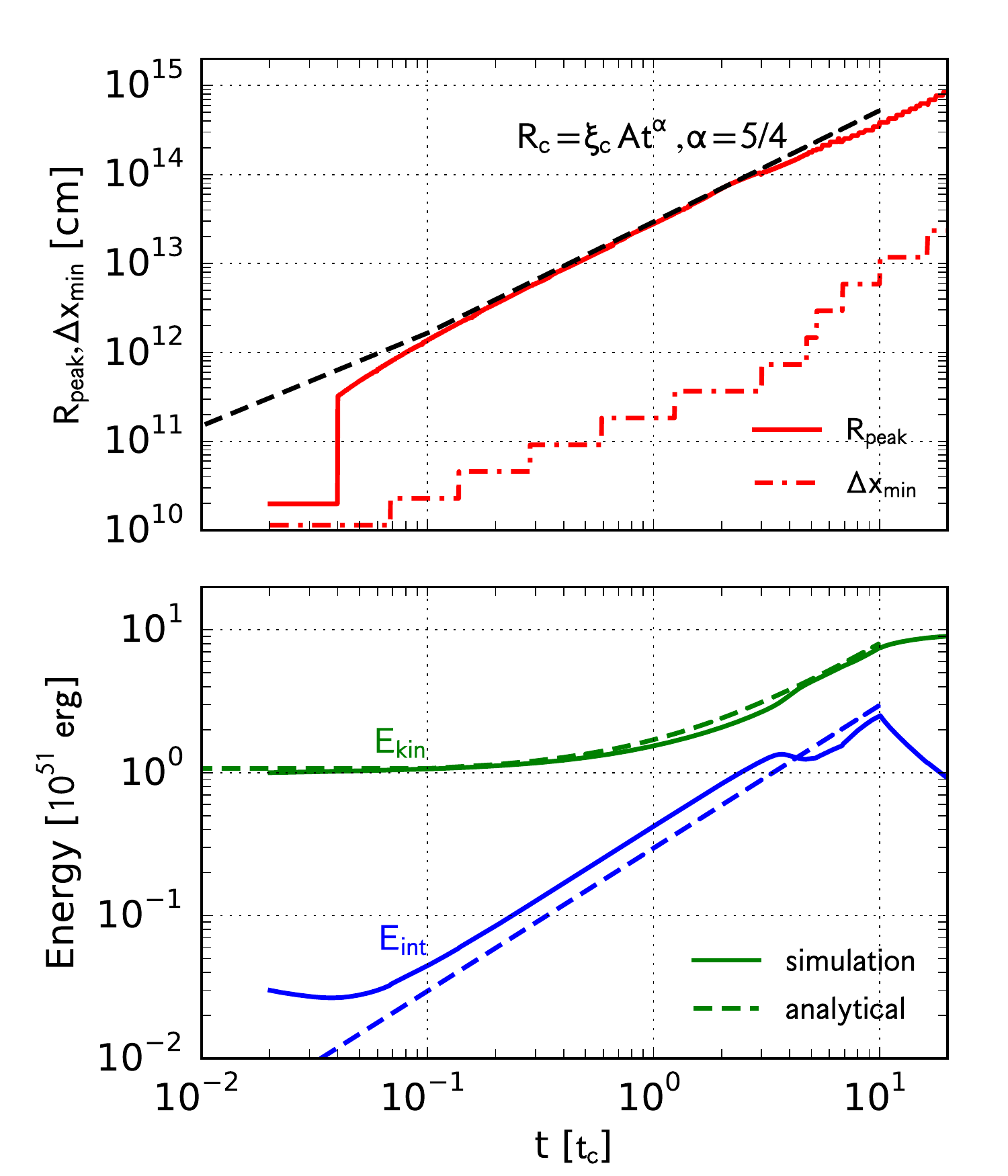}
\caption{Temporal evolutions of several physical variables. 
In the upper panel, the radius $R_\mathrm{peak}$ of the density peak and the minimum resolved length $\Delta x_\mathrm{min}$ are plotted as solid and dash-dotted lines. 
The temporal evolution of the radius $R_\mathrm{c}$ of the contact surface expected from the 1D self-similar solution is plotted as a dashed line and compared with the numerical result. 
In the lower panel, the kinetic and internal energies are plotted as solid lines. 
They are also compared with the expectations by the 1D self-similar solution. }
\label{fig:energy}
\end{center}
\end{figure}

We first focus on the results of model E52. 
The spatial distributions of the 4-velocity $\Gamma\beta$, the density $\rho$, and the pressure $p$, in the 2D slices around $y=0$ are shown in Figure \ref{fig:breakout}. 
The panels in Figure \ref{fig:breakout} represent snapshots of the simulation at $t/t_\mathrm{c}=3.0$, $4.0$. $4.5$, and $5.0$, during which the forward shock emerges from the surface of the ejecta. 
The dynamical evolution of the ejecta in model E52 is qualitatively similar to that of our previous 2D cylindrical simulation (\hyperlink{sm17}{SM17}) and calculations by other authors \citep{2016ApJ...832...73C,2017ApJ...845..139B}. 

At the beginning of the evolution, a quasi-spherical region with high pressure (hot bubble) appears around the center of the ejecta due to the energy injection. 
The forward shock driven by the pressure of the hot bubble propagates in the supernova ejecta. 
When the forward shock front is still in the inner part of the supernova ejecta, the ram pressure exerting on the post-shock gas is sufficiently large to contain the thermal pressure of the hot bubble, keeping the hot bubble nearly spherical. 
However, once the forward shock front passes through the interface separating the inner and outer parts of the ejecta, $r=v_\mathrm{br}t$, the ram pressure of the outer ejecta can no longer contain the hot bubble, leading to an efficient acceleration of the forward shock. 
On the course of the hot bubble expansion, the Rayleigh-Taylor instability develops around the contact surface, mixing the injected gas with the supernova ejecta. 
The forward shock modified by the development of the Rayleigh-Taylor instability efficiently expands in the outermost layer of the supernova ejecta, resulting in the violent breakout of the shock into the surrounding space. 
This hot bubble breakout plays an important role in producing clumpy ejecta, which is distinguished from the 1D spherical picture of supernova ejecta with central energy injection. 

In our 3D simulation, the presence of the hot bubble and its breakout are successfully reproduced, as is seen in Figure \ref{fig:breakout}. 
In our previous 2D cylindrical simulation, shocked region and the density distribution exhibit global bipolar structure despite the energy injection in a spherical manner. 
As we have noted in \hyperlink{sm17}{SM17}, this is an artificial effect due to the assumed axisymmetry in the 2D cylindrical simulation. 
The 3D results presented in Figure \ref{fig:breakout} do not suffer from such an artificial effect and exhibit globally spherical ejecta after the hot bubble breakout. 
However, one caveat should be noted. 
In the initial stage of the development of the Rayleigh-Taylor instability, the shock front exhibits an artificial symmetry. 
In this simulation, numerical errors arising from the 3D cartesian grid structure predominantly contribute to the seed of the instability. 
Therefore, the angular distributions of hydrodynamic variables should be treated carefully. 
This issue could be remedied by simulations with much higher resolution.
As we shall see below, however, the radial distributions of hydrodynamic variables after the hot bubble breakout can be understood in the same analytical way as the 2D cylindrical study, and thus our conclusions are unlikely affected by this numerical issue.

Figure \ref{fig:energy} shows the growth of the hot bubble in a more quantitative way. 
We identify the numerical cell with the highest density (referred to as the density peak) in the computational domain and plot the distance $R_\mathrm{peak}$ of the density peak from the center as a function of time in the upper panel of Figure \ref{fig:energy}. 
The growth of $R_\mathrm{peak}$ is expected to follow that of the contact surface separating the shocked injected gas and the surrounding ejecta. 
Its temporal evolution, which is shown as a solid line, is compared with a self-similar solution, which has been developed for SN remnants with a pulsar wind nebula in its center \citep{1984ApJ...280..797C,1992ApJ...395..540C,1998ApJ...499..282J}. 
The radius $R_\mathrm{c}$ of the contact surface is given by
\begin{equation}
R_\mathrm{c}=\xi_\mathrm{c}At^\alpha,
\label{eq:Rc}
\end{equation}
with $\alpha=5/4$ and $\xi_\mathrm{c}=0.9849$ for the parameter set adopted in this calculation. 
The constant $A$ is given by
\begin{equation}
A=\left\{
\frac{3(\gamma-1)(2-\gamma)L}{\alpha^2\gamma\left[1+3\alpha(\gamma-1)\right]
\xi_\mathrm{c}^3\eta_\mathrm{c}(v_\mathrm{br})^{\delta-3}f_3M_\mathrm{ej}}
\right\}^{1/(5-\delta)},
\label{eq:coeff_A}
\end{equation}
with $\eta_\mathrm{c}=0.1706$ (see, \hyperlink{sm17}{SM17} for the derivation of the self-similar solution). 
In the upper panel of Figure \ref{fig:energy}, we also plot the temporal evolution of the minimum resolved length $\Delta x_\mathrm{min}$. 
In the lower panel of Figure \ref{fig:energy}, we plot the kinetic and internal energies, $E_\mathrm{kin}$ and $E_\mathrm{int}$, as a function of time. 
From the self-similar solution, they are expected to grow as follows,
\begin{equation}
E_\mathrm{kin}=E_\mathrm{sn}+\frac{(1+3\alpha)(\gamma-1)}{1+3\alpha(\gamma-1)}Lt,
\label{eq:Ekin}
\end{equation}
for the kinetic energy, and
\begin{equation}
E_\mathrm{int}=\frac{(2-\gamma)}{1+3\alpha(\gamma-1)}Lt,
\label{eq:Eint}
\end{equation}
for the internal energy. 

The temporal evolution of $R_\mathrm{peak}$ agrees with that of $R_\mathrm{c}$, Equation (\ref{eq:Rc}), up to $t/t_\mathrm{c}\simeq 4.0$. 
The growth of the kinetic and internal energies are also well described by Equations (\ref{eq:Ekin}) and (\ref{eq:Eint}) in the same time interval. 
This means that the dynamical evolution of the ejecta immediately approaches the self-similar solution, thereby suggesting that the results of the simulations are less dependent on the value of the initial time $t_\mathrm{0}$ and the initial condition. 
The radius $R_\mathrm{peak}$ starts deviating from the self-similar solution after $t/t_\mathrm{c}\simeq 3.0$--$4.0$. 
This time corresponds to the breakout of the hot bubble from the outermost layer of the ejecta (see, Figure \ref{fig:breakout}). 
From the analytical considerations, the breakout is expected to occur when the forward shock driven by the hot bubble reaches the interface separating the inner and the outer parts of the ejecta, at $r=v_\mathrm{br}t$. 
The self-similar solution predicts that this occurs at $t=f_\mathrm{br}t_\mathrm{c}$, where the numerical factor is given by
\begin{equation}
f_\mathrm{br}=\frac{2\alpha^2\gamma\left[1+3\alpha(\gamma-1)\right]\xi_\mathrm{c}^3\eta_\mathrm{c}f_5}{3(\gamma-1)(2-\gamma)}=5.1, 
\end{equation}
for the adopted parameter set. 
The hot bubble breakout occurs slightly earlier in the numerical simulation than the prediction by the 1D self-similar solution. 
This is probably due to the shock front modified by the Rayleigh-Taylor instability. 
Because of the non-linear development of the Rayleigh-Taylor instability, shocks propagating into some radial directions can ``overshoot'' with respect to the average radius of the shock front. 
Such overshooting components emerge from the inner ejecta earlier than shocks along other directions, which makes the hot bubble breakout  happening earlier. 
This qualitatively explains the earlier hot bubble breakout. 
However, the development of the Rayleigh-Taylor instability before the hot bubble breakout is highly non-linear, which makes it difficult to predict the exact onset time of the hot bubble breakout.

\subsubsection{Density structure}
\begin{figure*}
\begin{center}
\includegraphics[scale=1.1,bb= 0 0 446 268]{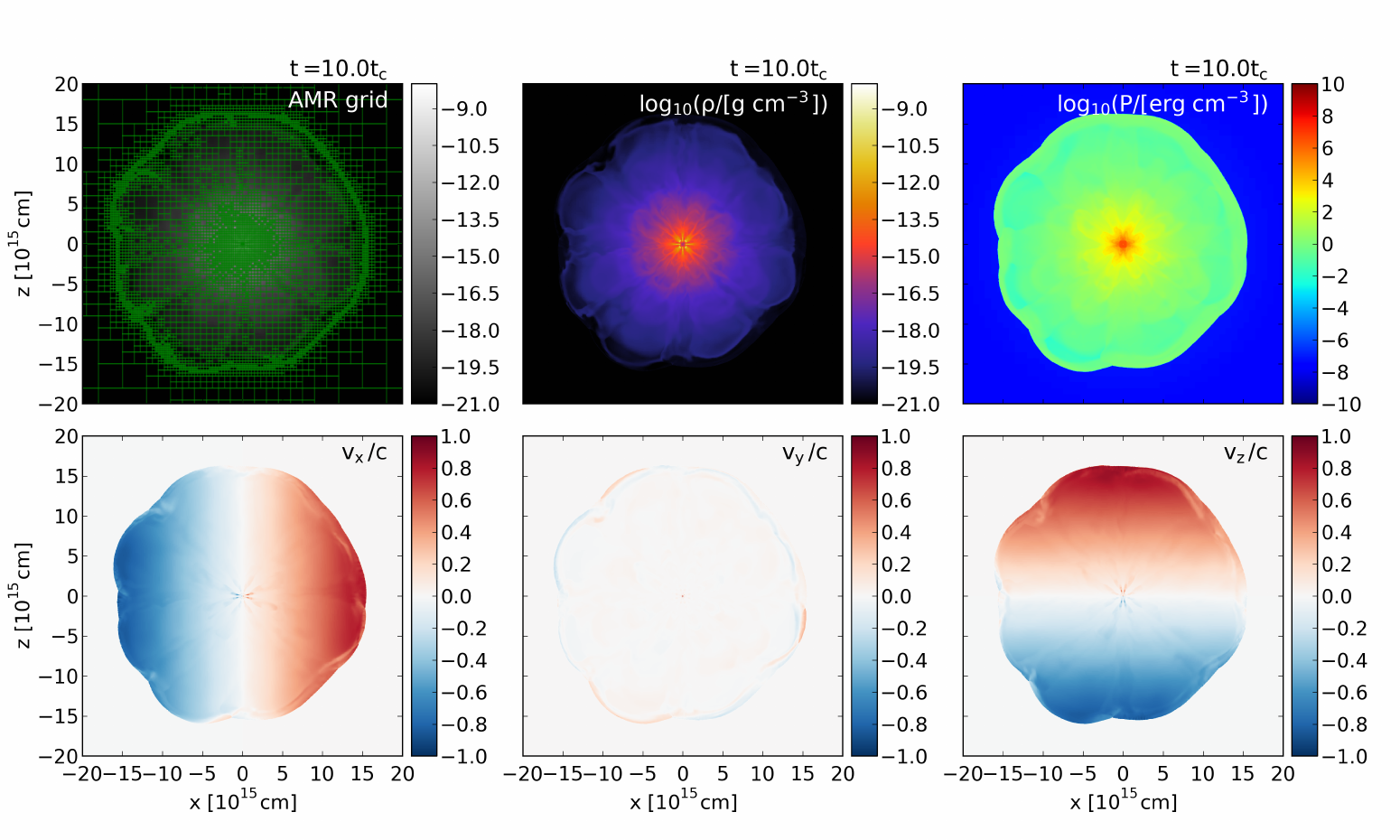}
\caption{Spatial distributions of physical variables at $t/t_\mathrm{c}=10.0$. 
The upper left, center, and right panels represent the AMR grid structure, density, and pressure, while lower panels represent the three components of the velocity. 
The supernova ejecta is already overwhelmed by the forward shock driven by the hot bubble, producing fast ejecta traveling at velocities close to the speed of light.}
\label{fig:snap_t100}
\end{center}
\end{figure*}

Figure \ref{fig:snap_t100} shows the density distribution at the end of the simulation, $t/t_\mathrm{c}=10$. 
The hot bubble breakout and a bunch of relativistic outflows penetrating the ejecta realize patchy density structure. 
The outermost layer of the supernova ejecta is accelerated by shocks driven by the hot bubble and their velocities are close to the speed of light. 
Since the central energy injection is terminated at $t/t_\mathrm{c}=10.02$, the outer part of the ejecta is already expanding almost freely. 

\begin{figure}
\begin{center}
\includegraphics[scale=0.55,bb= 0 0 453 566]{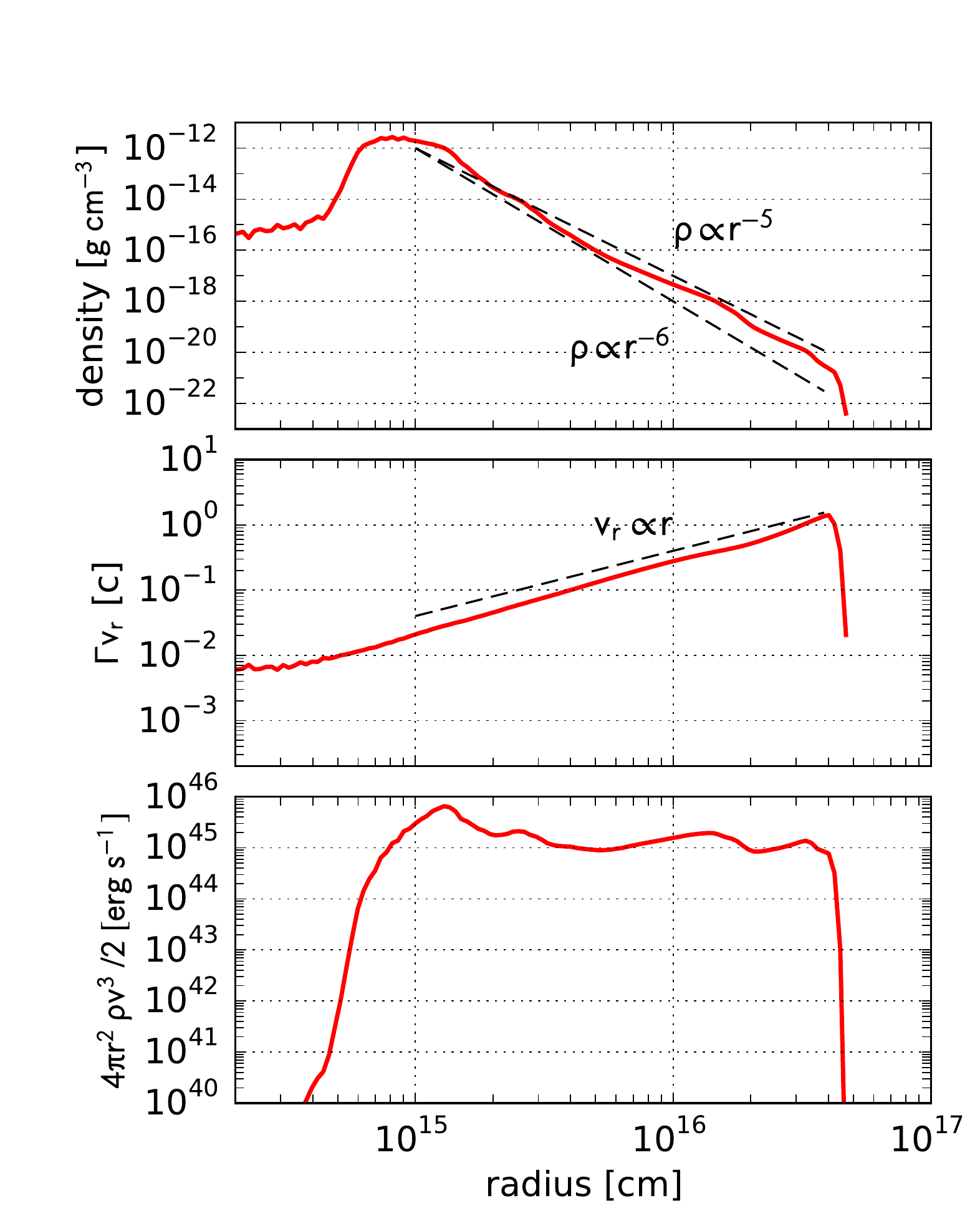}
\caption{Averaged radial distributions of the density, the radial four-velocity, and the kinetic luminosity at $t/t_\mathrm{c}=20$. }
\label{fig:radial}
\end{center}
\end{figure}

Despite the patchy density structure, physical variables exhibit simple global distributions. 
We obtain the radial profile of a physical variable $q$ by dividing the numerical domain into successive concentric shells with a small width $\Delta r$ and averaging the variable over the shell,
\begin{equation}
q(r)=\frac{1}{\Delta V}\int^{r+\Delta r/2}_{r-\Delta r/2}r^2dr\int_{4\pi} d\Omega q(x,y,z),
\end{equation}
where $\Delta V=4\pi r^2\Delta r$ is the volume of the concentric shell and $\Omega$ is the solid angle. 
In Figure \ref{fig:radial}, we plot the radial profiles of the density, the radial velocity $v_r$, and the kinetic luminosity at $t/t_\mathrm{c}=20$. 
The kinetic luminosity at $r$ is defined as the energy going through the surface area $4\pi r^2$ at $r$ per unit time and is calculated by the product of the non-relativistic kinetic energy flux $\rho v_r^3/2$ and the surface area $4\pi r^2$. 

The radial velocity profile follows $v_r\propto r$, which is naturally expected for freely expanding ejecta. 
The kinetic luminosity shown in the bottom panel of Figure \ref{fig:radial} exhibits a flat profile extending out to layers traveling at relativistic speeds, $\Gamma v_r\sim c$. 
This is understood as a result of the central energy injection at a constant rate and the efficient energy transport throughout the ejecta. 
As discussed in \hyperlink{sm17}{SM17}, the channel flows emanating the ejecta play a role in transporting the injected energy and depositing it into the outer layers of the ejecta, resulting in a flat kinetic energy distribution. 
This is a distinguishing feature of the multi-dimensional picture of supernova ejecta with a long-lived central energy source. 
In spherically symmetric models, the transport of the injected energy is realized by the spherical forward shock driven by the pressure of the hot bubble, ending up as a spherical shell made up of the shocked ejecta \citep{2010ApJ...717..245K}. 

\hyperlink{sm17}{SM17} have considered the radial density profile realized when freely expanding spherical ejecta follow a flat kinetic energy distribution, and analytically derived a single power-law density profile, $\rho\propto t^{-3}v_r^{-n}$ or $\rho\propto t^{n-3}r^{-n}$, with an exponent in the range of $5\leq n\leq 6$. 
In the top panel of Figure \ref{fig:radial}, we compare the radial density profile obtained by our hydrodynamic simulation with single power-law functions with exponents $-5$ and $-6$. 
The comparison shows a good agreement, demonstrating the validity of the analytical consideration. 

\subsection{Model E51}
\subsubsection{Dynamical evolution of ejecta with moderate energy injection}
\begin{figure*}
\begin{center}
\includegraphics[scale=1.2,bb= 0 0 446 502]{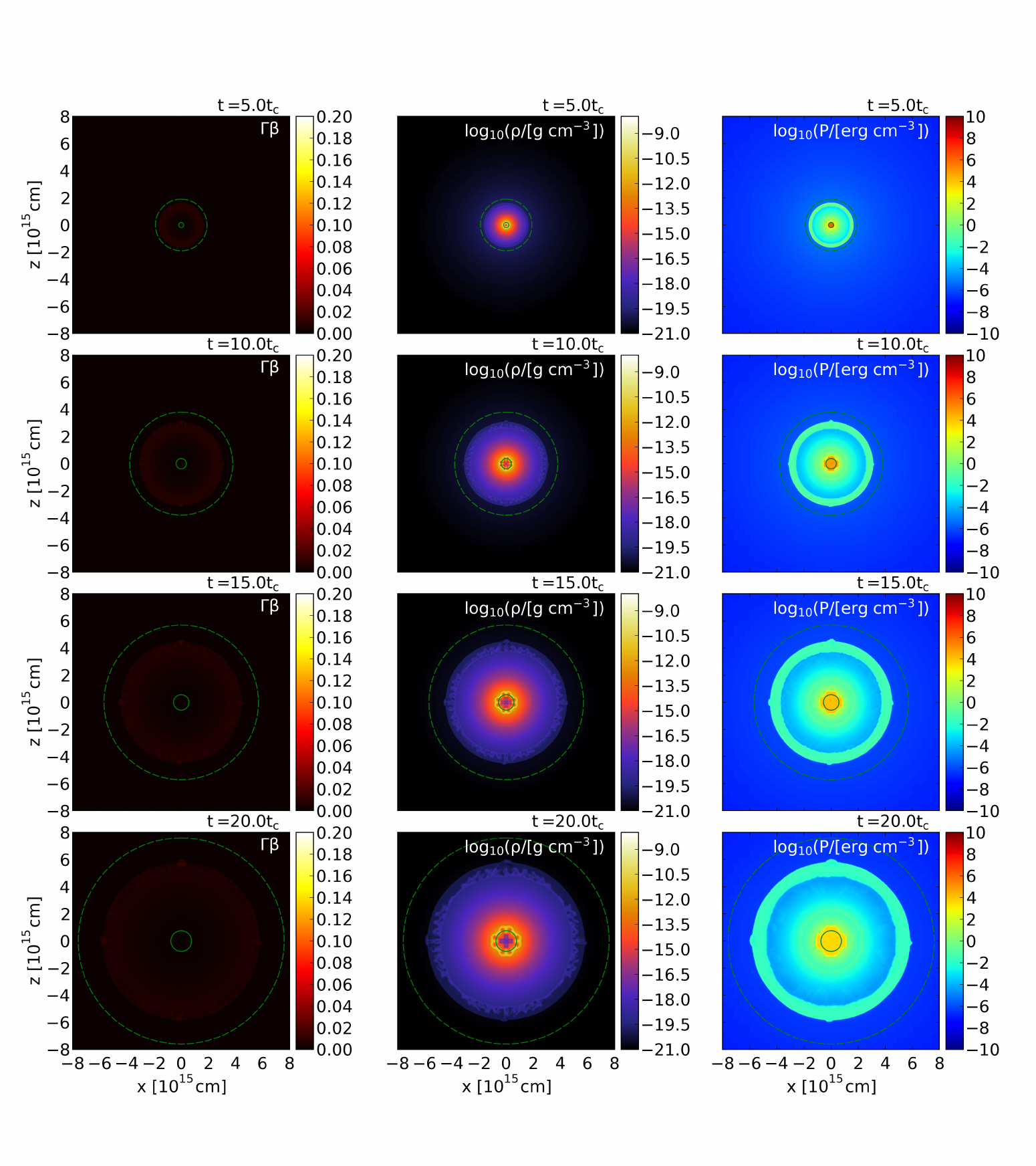}
\caption{Spatial distributions of physical variables at $t/t_\mathrm{c}=5.0$, $10.0$, $15.0$, and $20.0$ (from top to bottom) for model E51. }
\label{fig:E51}
\end{center}
\end{figure*}

\begin{figure}
\begin{center}
\includegraphics[scale=0.55,bb= 0 0 432 504]{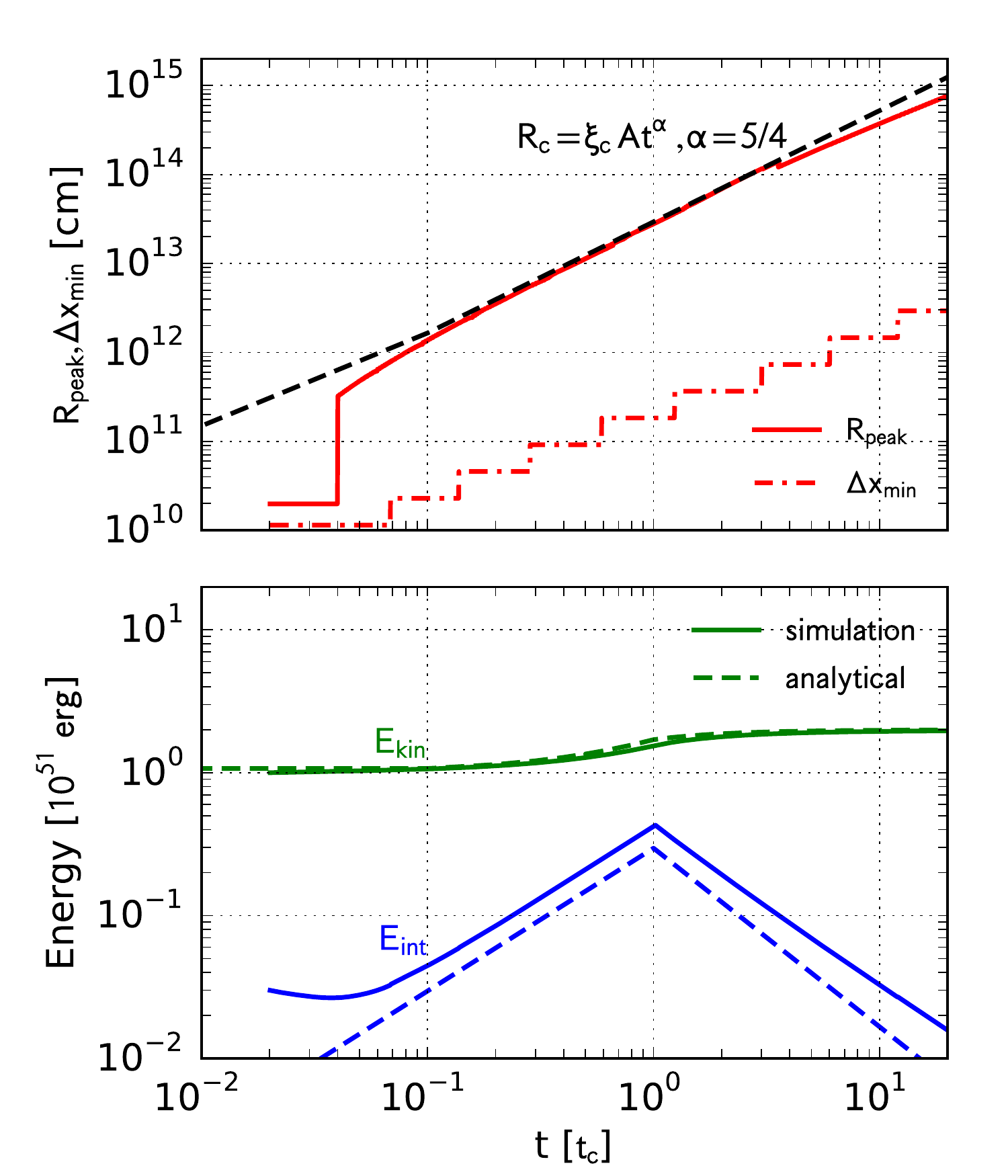}
\caption{Same as Figure \ref{fig:energy}, but for model E51. }
\label{fig:energy_E51}
\end{center}
\end{figure}

The dynamical evolution of model E51 shows a marked difference from its more energetic counterpart. 
Figure \ref{fig:E51} shows spatial distributions of physical variables realized in model E51 after the central energy injection is terminated. 
Although the hot bubble can be seen in the pressure distribution in Figure \ref{fig:E51}, it is still confined by the unshocked ejecta in a quasi-spherical manner. 
Even at the end of the simulation, $t/t_\mathrm{c}=20.0$ (the bottom panels of Figure \ref{fig:E51}), the forward shock front is far from the outermost layer of the ejecta. 
This is clearly due to the smaller injected energy than model E52. 

This outcome can be understood again by the self-similar solution described by Equations (\ref{eq:Rc}) and (\ref{eq:coeff_A}). 
In fact, the total amount of the injected energy $E_\mathrm{inj}$ relative to the initial kinetic energy $E_\mathrm{sn}$ of the supernova ejecta plays a critical role in determining whether the forward shock reaches the interface before the energy injection is terminated. 
When the energy injection still continues after the forward shock reaches the interface, the continuously supplied energy drives the violent breakout of the hot bubble as we have described in Section \ref{sec:E52_model}. 
However, without the on-going energy supply, the hot bubble appears to be stuck in the ejecta even after the forward shock reaches the interface. 
For the adopted density profile, the threshold energy of $5.1E_\mathrm{sn}\simeq 5\times 10^{51}$ erg (see, \hyperlink{sm17}{SM17}) is between $10^{51}$ and $10^{52}$ erg, which reasonably explains the different dynamical evolutions of models E52 and E51. 
Figure \ref{fig:energy_E51} presents the temporal evolutions of $R_\mathrm{peak}$, $\Delta x_\mathrm{min}$, and the kinetic and the internal energies. 
In contrast to model E52, the termination of the central energy injection at $t/t_\mathrm{c}= 1.0$ makes the internal energy suddenly decline at this time.

\subsubsection{Self-similar expansion of the hot bubble}\label{sec:expansion_of_stacking_bubble}
\begin{figure}
\begin{center}
\includegraphics[scale=0.55,bb= 0 0 453 566]{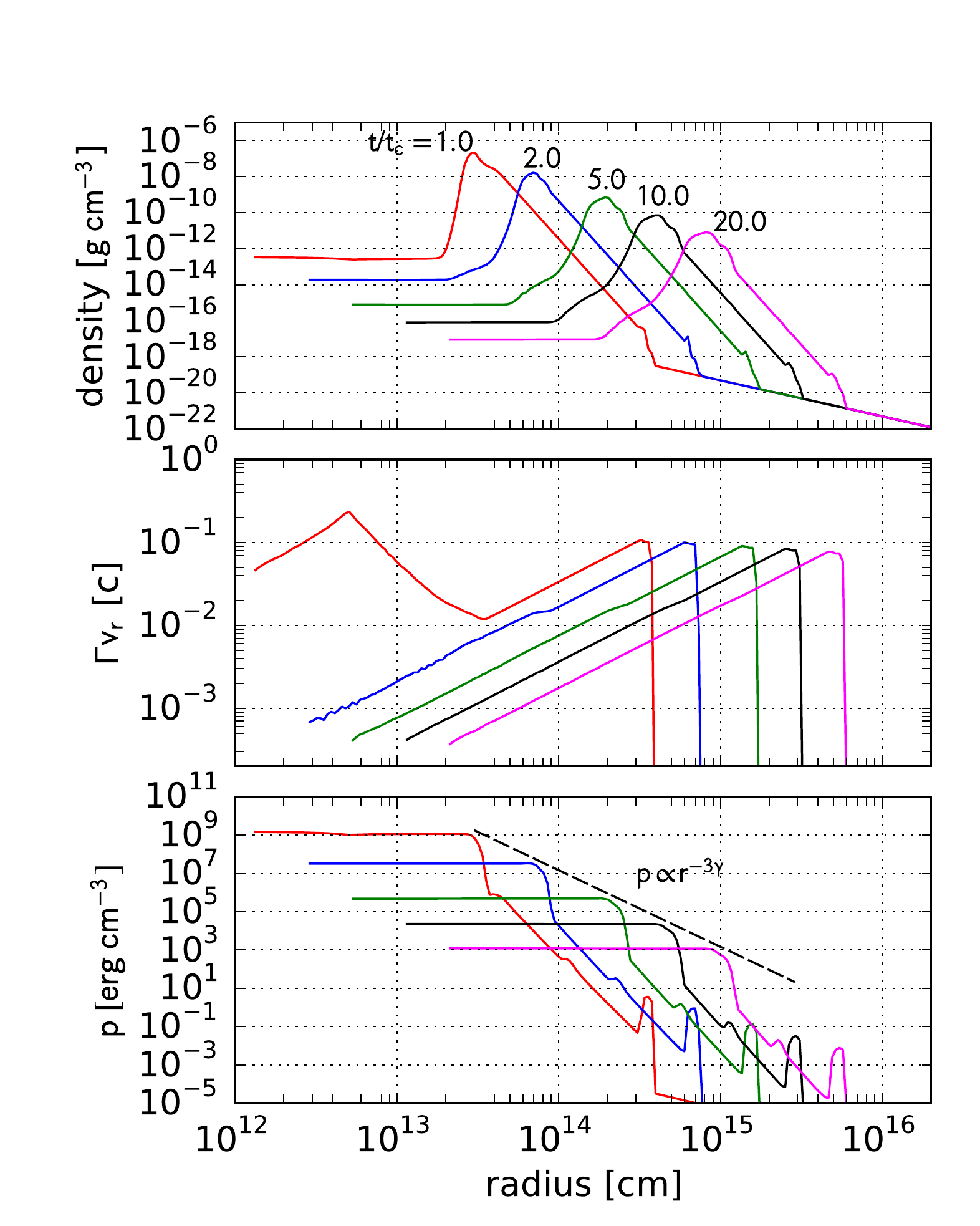}
\caption{Averaged radial distributions of the density, the radial four-velocity, and the pressure for model E51. 
In each panel, radial profiles at $t/t_\mathrm{c}=1.0$, $2.0$, $5.0$, $10.0$, $20.0$ are plotted from left to right. 
In the bottom panel, the scaling relation $p\propto r^{-3\gamma}$ is also plotted as a dashed line.}
\label{fig:radial_E51}
\end{center}
\end{figure}

\begin{figure}
\begin{center}
\includegraphics[scale=0.55,bb= 0 0 453 566]{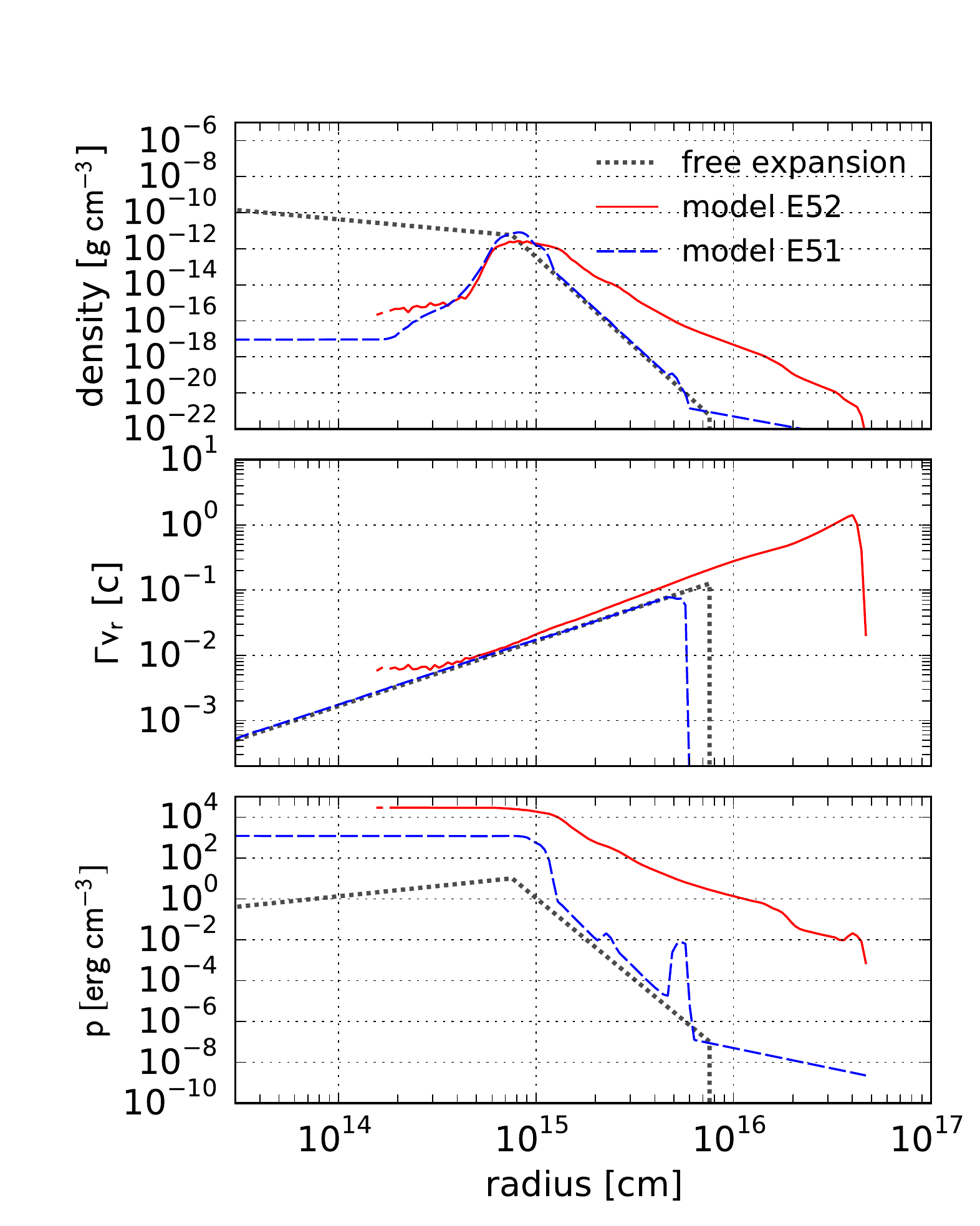}
\caption{Comparison of the radial profiles of the density (top), the radial 4-velocity (middle), and the pressure (bottom) at $t=20t_\mathrm{c}$. 
The solid and dashed lines represent profiles for models E52 and E51. 
In each panel, the distributions of freely expanding ejecta without additional energy injection are also plotted for comparison. }
\label{fig:radial_comparison}
\end{center}
\end{figure}

When the injected energy is insufficient to realize the hot bubble breakout, the hot bubble is confined in the ejecta even after the forward shock front reaches the outer part of the ejecta with the steep density gradient. 
Figure \ref{fig:radial_E51} shows the radial distributions of the density, the radial four-velocity and the pressure at $t/t_\mathrm{c}=1.0$, $2.0$, $5.0$, $10.0$, and $20.0$ for model E51. 
The radial profiles are composed of the unshocked ejecta and the embedded hot bubble. 
As is seen in the radial profiles of the pressure, the hot bubble is clearly separated from the unshocked ejecta by the forward shock front. 
The density profiles exhibit peaks around the interface between the hot bubble and the surrounding ejecta. 
One of the remarkable features is the almost uniform pressure distribution within the hot bubble. 
This suggests that the time scale of the shock propagation into the surrounding ejecta is sufficiently longer than the sound crossing time within the hot bubble, i.e., the nearly entire region of the hot bubble is causally connected. 
Therefore, the pressure equilibrium is achieved inside the hot bubble. 

Furthermore, the profiles of physical variables at different epochs are similar to each other, indicating that the hot bubble and the surrounding ejecta evolve in a self-similar way, although the contact surface has already suffered from the development of the Rayleigh-Taylor instability. 
This suggests that the evolution is described by another self-similar solution, which corresponds to a point-like explosion in a spherical expanding medium. 
In Appendix \ref{sec:self-similar}, we consider the self-similar expansion of a spherical hot region in a freely expanding medium with a power-law radial density profile, by assuming that the pressure equilibrium is realized in the entire shocked region. 
In deriving the self-similar expansion law, we have assumed that the hot bubble expands in an adiabatic way (equation \ref{eq:p_b}). 
We found that; (i) in a medium with a shallow radial density slope, $-d\ln \rho/d\ln r\leq 3\gamma+2$, the hot bubble is confined in the ejecta and passively expands according to the freely expanding ambient gas, and
(ii) in a medium with a steep radial density slope, $-d\ln \rho/d\ln r>3\gamma+2$, the forward shock expands outwardly with respective to the expanding ambient gas. 
This finding suggests that the hot bubble is stuck in the inner ejecta when its outer radius does not reach the interface separating the inner and outer ejecta, $r=v_\mathrm{br}t$. 
On the other hand, once the hot bubble emerges from the interface, the latter self-similar expansion law for steep radial density slopes can apply. 

In fact, the injected energy relative to the initial ejecta kinetic energy, $E_\mathrm{inj}/E_\mathrm{sn}$, again plays a role in determining which case should be realized. 
In order for the outer radius of the hot bubble to catch up with the interface, the forward shock velocity $v_\mathrm{fs}$ relative to the velocity of the ambient ejecta should be larger than or the same order of magnitude as the break velocity $v_\mathrm{br}$. 
When the hot bubble sweeps a mass $M_\mathrm{sw}$, the forward shock velocity can roughly be estimated to be $v_\mathrm{fs}\sim (2E_\mathrm{inj}/M_\mathrm{sw})^{1/2}$. 
Setting the swept mass to be the mass of the inner ejecta, 
\begin{equation}
M_\mathrm{sw}\simeq\frac{m-3}{m-\delta}M_\mathrm{ej},
\end{equation}
the forward shock velocity is found to be
\begin{equation}
v_\mathrm{fs}\simeq\left[\frac{2(m-\delta)E_\mathrm{inj}}{(m-3)M_\mathrm{ej}}\right]^{1/2},
\end{equation}
when it reaches the interface. 
For $E_\mathrm{inj}\simeq E_\mathrm{sn}$, which is of particular interest here, this forward shock velocity is the same order of magnitude as $v_\mathrm{br}$. 
Therefore, in this particular model E51, the hot bubble can reach the transition layer $r=v_\mathrm{br}t$. 
As is shown in Figures \ref{fig:E51} and \ref{fig:radial_E51}, the hot bubble certainly reaches the outer ejecta at later epochs. 
In addition, in the bottom panel of Figure \ref{fig:radial_E51}, we plot the dashed line representing the scaling relation $p\propto r^{-3\gamma}$ expected for the adiabatic expansion, and confirm the agreement with the declining hot bubble pressure. 
Therefore, the late-time behavior of the hot bubble in the numerical simulation is well explained by the latter case. 
On the other hand, for a sufficiently small injected energy $E_\mathrm{inj}\ll E_\mathrm{sn}$, the hot bubble would never reach the interface and end up being dissolved in the inner ejecta. 

To summarize, the dynamical evolution of the hot bubble in model E51 first follows the self-similar solution discussed in Section \ref{sec:hot_bubble_expansion_and_breakout}, and then makes a transition to the other self-similar evolution discussed in Appendix \ref{sec:self-similar}.

Due to the terminated energy injection, the internal energy of the ejecta decreases after the linear increase described by Equation (\ref{eq:Eint}). 
While the forward shock is propagating in the outer part of the ejecta, the self-similar expansion law expects the internal energy decreasing as $E_\mathrm{int}\propto t^{-5/4}$. 
Thus, we obtain the following temporal evolution of the internal energy by connecting the two different self-similar laws,
\begin{equation}
E_\mathrm{int}(t)=\left\{
\begin{array}{cl}
\frac{(2-\gamma)}{1+3\alpha(\gamma-1)}L(t-t_0)&\mathrm{for}\ t\leq t_\mathrm{term},\\
\frac{(2-\gamma)}{1+3\alpha(\gamma-1)}E_\mathrm{inj}\left(\frac{t}{t_\mathrm{term}}\right)^{-5/4}
&\ \mathrm{for}\ t_\mathrm{term}<t.
\end{array}\right.
\end{equation}
Accordingly, the kinetic energy evolves as,
\begin{equation}
E_\mathrm{kin}(t)=E_\mathrm{sn}+L(t-t_\mathrm{0})-E_\mathrm{int}(t).
\end{equation}
In the bottom panel of Figure \ref{fig:energy_E51}, we compare the temporal evolutions of the kinetic and internal energies in the numerical simulation with the analytical evaluation, which shows an overall agreement. 

We estimate the time when the forward shock front catches up with the outermost layer of the ejecta. 
At the end of the simulation, $t/t_\mathrm{c}=20$, the forward shock is located around $R_\mathrm{fs,f}=10^{15}$ cm, which continues to grow as $R_\mathrm{fs}\propto t^{5/4}$. 
Thus, the forward shock front reaches the surface of the ejecta when the following condition,
\begin{equation}
R_\mathrm{fs,f}\left(\frac{t}{20t_\mathrm{c}}\right)^{5/4}=v_\mathrm{max}t,
\end{equation}
is satisfied, where we have neglected the interaction between the outermost layer and the surrounding medium for simplicity. 
For the maximum velocity of $v_\mathrm{max}\simeq 3.8\times10^9$ cm s$^{-1}$, the expected ``touch-down'' time is found to be
\begin{equation}
t=\frac{v_\mathrm{max}^4(20t_\mathrm{c})^5}{R_\mathrm{fs,f}^4}\simeq 2\times10^2\ \mathrm{yrs}.
\end{equation}
The outermost layer of the ejecta is eventually decelerated by the collision with the ambient medium, which makes the touch-down happen earlier to some extent. 
Nevertheless, this time scale is clearly much longer than the diffusion time scale for thermal photons in the ejecta. 
We also note that the diffusion time scale shorter than the touch-down time violates the assumption of the adiabatic expansion in deriving the self-similar expansion law. 
When approximately taking into account the effective radiative loss, where the adiabatic exponent of the post-shock gas is approximately close to unity, the hot bubble is stuck in the ejecta as we discuss in Appendix \ref{sec:self-similar}. 

Finally, in Figure \ref{fig:radial_comparison}, we compare the radial distributions of the density, the radial 4-velocity, and the pressure at $t=20t_\mathrm{c}$ for models E52 and E51. 
We again see the differences between the ejecta with and without the hot bubble breakout. 
We also plot the radial distributions of these variables for the freely expanding ejecta without the central energy injection nor the interaction with the ambient medium. 
For model E52, the ejecta is clearly accelerated and thus the outer layer of the ejecta is well ahead of the free expansion case. 
For model E51, while the inner part of the ejecta is significantly affected by the energy injection, the outer part exhibits similar radial profiles to the freely expanding case. 

\section{Discussion}\label{sec:discussions}
Through the 3D hydrodynamical simulations, we have shown that supernova ejecta with a long-lived central energy source follow different dynamical evolutions depending on the total amount of the injected energy. 
In the following, we summarize the dynamical evolution of supernova ejecta with a central energy source and discuss their observational implications.

\subsection{Supernova ejecta with hot bubble breakout}\label{sec:supernova_ejecta_with_hot_bubble_breakout}
First, we consider the evolution of supernova ejecta with a sufficiently large injection energy for the hot bubble breakout. 
Supernova ejecta with relatively large kinetic energies would have an advantage in explaining some energetic and/or bright supernovae. 
Several energetic SNe Ic-BL associated with GRBs, e.g., SNe 1998bw \citep{1998Natur.395..663K,1998Natur.395..670G}, 2006aj \citep{2006Natur.442.1008C,2006Natur.442.1018M,2006ApJ...645L..21M,2006Natur.442.1011P,2006Natur.442.1014S}, and 2010bh \citep{2011MNRAS.411.2792S,2011ApJ...740...41C,2012ApJ...753...67B,2012A&A...539A..76O} are believed to be driven by a powerful central engine. 
Even without any associated gamma-ray signal explicitly manifesting central engine activities, radio-bright SNe Ic-BL sharing similar properties with GRB-SNe, such as SN 2009bb and 2012ap, are also suspected to be an explosion driven by a central engine \citep{2010Natur.463..513S,2015ApJ...799...51M,2015ApJ...805..187C}. 
In addition, some authors pointed out that the energetic type Ib SN 2012au also share common emission properties with energetic SNe Ic-BL \citep{2013ApJ...770L..38M,2013ApJ...772L..17T,2018ApJ...864L..36M}. 
The observational properties of these energetic SNe are worth comparing with the multi-dimensional picture of supernova ejecta with a powerful central energy source revealed by the present simulation.

\subsubsection{Photometric and spectral evolution}
Our 3D hydrodynamic simulation turns out to be qualitatively similar to the 2D cylindrical counterpart, confirming the multi-dimensional picture of supernova ejecta with a powerful central energy source. 
As has already been suggested by several work (\citealt{2016ApJ...832...73C}; \hyperlink{sm17}{SM17}), one of the important outcomes of the hot bubble breakout is the efficient mixing of materials out to the surface of the ejecta. 
The supernova ejecta penetrated by channel flows likely show spatially inhomogeneous or clumpy distributions of elements having synthesized in the steady and explosive burning stages prior to the energy injection. 
Therefore, heavy elements supposed to be embedded in deeper layers for normal supernova ejecta could be found in both inner and outer layers for supernova ejecta with hot bubble breakout. 
These mixed heavy elements would contribute to opacities in outer layers of the ejecta, affecting the photometric and spectroscopic properties of SNe. 
Radioactive nickel mixed into outer layers of SN ejecta is also known to affect light curves of SNe. 
Recent systematic analysis of SNe Ic-BL from the PTF \citep{2019A&A...621A..71T} found that all of them require nickel mixing to some extent in order to explain their light curves. 
The spectroscopic observations of GRB 171205A/SN 2017iuk at unprecedentedly early epochs \citep[][]{2019Natur.565..324I} recently revealed the presence of heavy metals in the fast ejecta component preceding the SN ejecta. 
These are in agreement with the elemental mixing due to a long-lived central engine.

Another important consequence of the hot bubble breakout is relatively flat radial density distributions. 
The radial density structure is of fundamental importance in spectral formation. 
As we have seen in Section \ref{sec:E52_model}, the radial density profile of the ejecta is well described by a power-law function with an exponent between $-5$ and $-6$. 
This density profile is remarkably shallow compared with outer envelopes of normal SNe \citep{1989ApJ...341..867C,1999ApJ...510..379M}. 
For the energetic SN Ic-BL 1997ef, \cite{2000ApJ...545..407M} introduced a flat outer density slope, $\rho\propto r^{-4}$, for better reproducing its spectral evolution, rather than the steeper one, $\rho\propto r^{-8}$, in the hypernova model of \cite{2000ApJ...534..660I}. 
This modification requires the total explosion energy of $1.75\times10^{52}$ erg for SN 1997ef, well exceeding the typical value of $10^{51}$ erg. 
Although the slope with the exponent $-4$ is even shallower than that found in our simulation, the ejecta structure with the flat density slope and the sufficiently large kinetic energy is in an overall agreement with the hydrodynamic model explored in this work. 
More recently, the spectral modelings of several SLSNe-I by \cite{2016MNRAS.458.3455M} have found that a steeper outer density slope, $\rho\propto r^{-7}$, is more appropriate, which is roughly consistent with the original density structure expected in a standard SN explosion. 
The different outer density structure may indicate the difference in the energy redistribution process in the two energetic populations of SNe Ic-BL and SLSNe-I.

In addition, the clumpy SN ejecta make it easier for high-energy photons and particles from the central energy source to penetrate through the ejecta \citep{2003ApJ...589..871A}. 
As a result, the energy deposition from the central compact object to the surrounding gas is realized in a more extended manner than well-structured spherical ejecta. 
Recently, \cite{2019A&A...621A.141D} carried out non-LTE spectral synthesis calculations of SNe with an extended energy injection region. 
He confirmed that the way of the energy injection surely affects light curves. 
The more extended the energy deposition is, the earlier thermal emission can escape from the ejecta, making them brighter in early epochs. 

The ionization states of different layers in the ejecta and the associated emission lines would also be affected. 
Indeed, in the recent modelings of the nebular spectra of some SLSNe-I by \cite{2017ApJ...835...13J}, they introduced clumpy ejecta in order to explain \ion{O}{1} recombination lines and other spectral features. 
\cite{2019A&A...621A.141D} also investigated the effect of clumpy ejecta \citep[see, also,][]{2018A&A...619A..30D}. 
The comparison of spectral synthesis calculations with some SLSNe-I spectra suggests that SLSNe-I require clumpy ejecta to some extent even at maximum light. 
The hot bubble breakout and the subsequent efficient mixing of the ejecta give one possible explanation for the clumpy ejecta.

\subsubsection{Non-thermal emission}
Another important feature of supernova ejecta with the hot bubble breakout is the presence of a mildly relativistic component in the ejecta. 
The channel flows deposit a relatively large amount of energies into a small fraction of the ejecta at outer layers, realizing a high energy-to-mass ratio at the outermost layer. 
It is widely recognized that supernova ejecta colliding with an ambient medium create the forward and reverse shocks, producing non-thermal particles \citep[see][for a review]{2017hsn..book..875C}. 
The fast ejecta can efficiently dissipate the kinetic energy in the presence of a dense ambient medium and give rise to bright non-thermal emission via synchrotron and inverse Compton processes. 
Some energetic SNe Ic-BL are indeed known as bright radio sources. 
In the context of radio-bright SNe, radial density profiles with a flat slope are likely to explain their bright radio emission. 
For example, \cite{2015ApJ...805..187C} suggested $\rho\propto r^{-6}$ for SN 20012ap. 
Recently, we have calculated radio and X-ray emission from supernova ejecta with power-law density profiles implied by the hot bubble breakout \citep{2018MNRAS.478..110S}. 
We found that ejecta with radial profiles of $\rho\propto r^{-5}$ or $\rho\propto r^{-6}$, could produce bright radio and X-ray emission with luminosities similar to those of radio-bright SNe Ic-BL, SN 1998bw and 2009bb. 

On the other hand, in spite of dedicated efforts to detect radio, X-ray, and gamma-ray emission from SLSNe-I, any plausible signal have not been detected up to a few years after the explosion (\citealt{2013ApJ...771..136L,2018ApJ...856...56C,2018ApJ...864...45M,2018A&A...611A..45R,2018ApJ...857...72H}, but see the case of SCP 06F6; \citealt{2009ApJ...697L.129G}). 
This may indicate that SLSNe-I do not accompany a high-velocity ejecta component, although it is a natural outcome of the violent hot bubble breakout. 
As we suggested in \cite{2018MNRAS.478..110S}, in SLSNe-I, most of the injected energy may be used for bright thermal emission rather than accelerating the forward shock, resulting in less violent or even the absence of the hot bubble breakout. 
However, the sample size is still limited and thus it is probably too early to conclude. 
Future survey and follow-up observations of SLSNe-I across wide wavelength ranges would figure out if most SNSNe-I lack radio and X-ray signals indicating central engine activities. 

The recent detection of a persistent radio source at the position consistent with the SLSN-I PTF10hgi \citep{2019ApJ...876L..10E} might indicate that the SLSN-I was actually powered by a central compact object and the radio emission from the wind nebula driven by the compact source is now unveiled $\sim 9$ years after its discovery. 
If the spatial coincidence would turn out to be true, this detection along with tight radio upper limits for other SLSNe-I in a few years might imply that radio emission can penetrate the surrounding gas only after it has become transparent to radio waves in $\sim10$ years, which is in fact consistent with theoretical expectations \citep{2018MNRAS.474..573O,2018MNRAS.481.2407M}.  
The wind nebula embedded in the SN ejecta is also expected to produce high-energy photons, which ionize the SN ejecta and eventually escape into interstellar space \citep[i.e., ionization breakout;][]{2014MNRAS.437..703M,2018MNRAS.481.2407M}. 
Since the clumpy density structure of the SN ejecta would make it easier for such high-energy photons and/or radio waves to escape, such electromagnetic signals may be observed earlier than the theoretical expectations based on spherical SN ejecta.  

\subsection{Supernova ejecta without hot bubble breakout}\label{sec:supernova_ejecta_without_hot_bubble_breakout}
In contrast, the density structure of the supernova ejecta with a moderately powerful energy source would be well stratified in the absence of the hot bubble breakout. 
The hot bubble is well confined by the ram pressure of the ejecta balancing the thermal pressure of the hot gas. 
Therefore, the internal energy of the hot bubble is eventually lost due to adiabatic expansion and radiative diffusion. 
It is the diffusion time throughout the ejecta that determines the dominant energy loss process. 
When the diffusion time scale is much longer than the time scale of the adiabatic expansion, the hot bubble eventually cools by exerting pressure on the surrounding gas and most energy would be shared among the ejecta as the kinetic energy. 
On the other hand, with the diffusion time scale shorter than the expansion time scale, the internal energy of the hot bubble can diffuse out through layers stratified on the hot bubble rather than being lost by adiabatic cooling. 
Specifically, the radiation front can overtake the shock front and then emerge from the outermost layer. 
As we have estimated in Section \ref{sec:expansion_of_stacking_bubble}, the time scale required for the forward shock to reach the outer edge of the ejecta is of the order of $100$ years. 
Thus, the diffusion time scale is clearly much shorter than this time scale, suggesting that a considerable fraction of the internal energy of the hot bubble would be lost via radiative diffusion. 

In this case, the resultant ejecta are composed of a high-density shell confining the hot bubble and well-stratified outer layers of the unshocked ejecta. 
The outer layers are kept hot due to diffusing photons from the hot bubble. 
These components would keep spherical symmetry to some extent as long as the energy injection and the original SN ejecta are roughly spherical. 
In other words, situations expected in 1D spherical calculations \citep[e.g.,][]{2010ApJ...717..245K} can apply. 
Therefore, bright thermal emission in optical wavelengths is expected, depending on the time scales of adiabatic cooling and radiative loss. 
The photospheric temperature is high and elements in the outer layers of the ejecta are highly ionized because of the power source. 
However, the expected density slope of the outermost layer is similar to normal stripped-envelope CCSNe, leading to spectral lines with moderate widths. 
As we have discussed in Section \ref{sec:supernova_ejecta_with_hot_bubble_breakout}, SNe with bright thermal emission but without bright radio and X-ray emission could be explained by this scenario.

\subsection{Two types of supernovae powered by central engines?}
The considerations in Sections \ref{sec:supernova_ejecta_with_hot_bubble_breakout} and \ref{sec:supernova_ejecta_without_hot_bubble_breakout} suggest that there may be two types of supernovae with central power sources, supernovae with relatively powerful and moderate additional energy sources. 
The former population would show bright optical emission and broad-line spectral features, and potentially be a bright source in radio, X-ray, and even gamma-ray. 
The latter population could also give rise to bright optical emission powered by a central energy source but without bright non-thermal emission. 

These findings can be related to the potential diversity of SLSNe-I. 
Whether or not SLSNe-I are made up of two or more sub-classes has been paid great attention. 
Investigating their observational properties, some authors have claimed that there are fast- and slow-evolving populations \citep[e.g.,][]{2013ApJ...770..128I,2017MNRAS.468.4642I}. 
\cite{2018ApJ...855....2Q} classified SLSNe-I in their PTF samples into PTF12dam-like and 2011ke-like objects, where the latter show smoother spectra around their optical maxima than the former. 
One simple interpretation of the systematic difference in their spectral appearance is that it reflects different ejecta structures. 
For a shallow radial density profile, a specific range of density corresponds to wider radius and velocity ranges than for a steep radial density profile, making absorption troughs broad. 
Therefore, the population of SLSNe-I with smooth spectra may indicate relatively flat density profiles. 

The different light curve evolution of SLSNe-I can also be a key to unveiling the potential diversity of SLSNe. 
Fast-evolving SLSNe-I appear to have smaller ejecta mass than the slowly-evolving population \citep{2015MNRAS.452.3869N}. 
Given that the ejecta mass is positively correlated with the explosion energy for normal stripped-envelop SNe \citep[e.g.,][]{2016MNRAS.457..328L}, ejecta with a smaller total mass are more likely to be affected by the central energy injection. 
In other words, even when the same injection energy is assumed, less massive ejecta (supposedly larger $E_\mathrm{inj}/E_\mathrm{sn}$) more likely experience the efficient mixing caused by the central energy injection.  
The 2011ke-like objects analyzed by \cite{2018ApJ...855....2Q} indeed exhibit fast evolving light curves, indicating smaller ejecta masses. 
Thus, the different spectral behavior may be explained by the different impacts of the central energy injection into the SN ejecta.

Unified scenarios for SLSNe-I and SN Ic-BL \citep[e.g.,][]{2015MNRAS.454.3311M,2016ApJ...818...94K} indicate that they share several common photometric and spectroscopic properties. 
In fact, there are some observational implications on the connection, such as, spectroscopic properties \citep{2010ApJ...724L..16P,2016ApJ...828L..18N,2017ApJ...835...13J,2017ApJ...845...85L}, and host galaxies \citep{2011ApJ...727...15N,2014ApJ...787..138L,2015MNRAS.449..917L,2015MNRAS.451L..65T,2016MNRAS.458...84A,2016ApJ...830...13P,2017MNRAS.470.3566C,2018MNRAS.473.1258S}. 
More recently, \cite{2019ApJ...872...90B} reported a transitional event, SN 2017dwh, which outshined as brightly in optical bands as SLSNe-I at early epochs, but showed a dramatic transition from an SLSNe-like blue spectrum to a red and broad-lined one similar to SN Ic-BL spectra. 
They also identified an absorption feature around $3200\mathrm{\AA}$, which they attributed to \ion{Co}{2}. 
This requires highly efficient mixing of iron-peak elements, which is usually embedded in the deep interior of the ejecta, to outer layers. 
A sufficiently powerful engine can potentially explain the highly bright emission and efficient mixing at the same time, as demonstrated by our simulations. 

\subsection{Final remarks}
Finally, we make some remarks on the simulation results and the scenario introduced above. 

The first remark is the resolution dependence of the numerical simulations. 
Even with the high spatial resolution realized by the AMR technique, the development of the hydrodynamic instability in our simulations show some artificial structure along the numerical grid. 
For example, as seen in Figure \ref{fig:breakout}, the Rayleigh-Taylor fingers are elongated along the $x$- and $z-$axes before the breakout. 
This is probably because of the grid structure. 
Therefore, care must be taken for the angular dependence of the column density or other hydrodynamic variables important for leakage of photons and high energy particles, and viewing angle effects. 
Nevertheless, the angle-averaged distributions, which are mainly discussed in this study, can be understood in analytical ways and therefore is robust.

We also note that our simulations do not take into account radiative transfer effects. 
For short diffusion time scales for photons in the ejecta, the energy loss via radiative diffusion must decrease the pressure of the hot bubble, probably making the breakout less violent or even suppressing it. 
This is expected when the diffusion time scale is comparable to the characteristic time scale of the energy injection, $E_\mathrm{inj}/L$. 
This effect should be investigated by radiation-hydrodynamic simulations appropriately incorporating radiative transfer effects, which we leave as a future study. 

\section{Conclusions}\label{sec:conclusions}
In this work, we performed 3D hydrodynamic simulations of supernova ejecta with a central energy source and presented their results. 
We performed simulations with the injected energy much larger than and comparable to the initial kinetic energy of the ejecta to reveal how the amount of the injected energy affects the dynamical evolution of the ejecta. 
The simulations with two different setups show a remarkable contrast, confirming the expectation that the total amount of the injected energy relative to the original explosion energy of the supernova is one of the most important parameters. 
The two qualitatively different models lead to different density structures, which can be probed by spectroscopic observations of extraordinary SNe. 
We point out that the engine-driven SN ejecta with different density structure can explain the diversity of energetic SNe. 
\\

\acknowledgements
We appreciate the anonymous referee for his/her constructive comments, which greatly helped us improve the manuscript. 
Numerical computations were carried out in part on the Cray XC50 at the Center for Computational Astrophysics, National Astronomical Observatory of Japan. 
A.S. acknowledges support by Japan Society for the Promotion of Science (JSPS) KAKENHI Grand Number JP19K14770. 
K.M. acknowledges support by JSPS KAKENHI Grant Numbers JP18H05223, JP18H04585, and JP17H02864. 

\appendix
\section{Self-similar expansion of a hot bubble in a power-law ejecta}\label{sec:self-similar}
In this section, we consider the expansion of a spherical hot region in freely expanding ejecta. 
We especially focus on the self-similar behavior of the forward shock and the pressure of the hot bubble so that the solution can be applied to the numerical result in Section \ref{sec:results}. 

We assume the following power-law density profile for the freely expanding ejecta corresponding to the outer envelope, 
\begin{equation}
\rho(t,r)=Bt^{m-3}r^{-m}.
\label{eq:rho_ej}
\end{equation}
The spherical hot region is located in the central region of the ejecta and is surrounded by the forward shock front propagating in the ejecta. 
The forward shock is located at $r=R_\mathrm{fs}(t)$ and evolves in a self-similar way,
\begin{equation}
R_\mathrm{fs}(t)=Ct^\lambda,
\end{equation}
where the exponent $\lambda$ is determined later. 
In order for the shock front to travel outward with respect to the freely expanding medium, the exponent $\lambda$ is required to be larger than unity, $\lambda>1$. 
The volume $V_\mathrm{b}$ of the hot bubble is thus given by,
\begin{equation}
V_\mathrm{b}=\frac{4\pi}{3}R_\mathrm{fs}^3=\frac{4\pi}{3}C^3t^{3\lambda}.
\end{equation}
The temporal evolution of the pre-shock density is obtained by inserting the forward shock radius into Equation (\ref{eq:rho_ej}),
\begin{equation}
\rho(t,R_\mathrm{fs})=BC^{-m}t^{-\lambda m+m-3}.
\end{equation}
The jump condition for a strong shock leads to the following post-shock pressure $p_\mathrm{f}$,
\begin{equation}
p_\mathrm{f}=\frac{2}{\gamma+1}\rho(t,R_\mathrm{fs})\left(\frac{{dR}_\mathrm{fs}}{dt}-\frac{R_\mathrm{fs}}{t}\right)^2=
\frac{2(\lambda-1)^2}{\gamma+1}BC^{2-m}t^{(2-m)\lambda+m-5}.
\label{eq:p_f}
\end{equation}

On the other hand, the internal energy $E_\mathrm{int}$ of the hot bubble evolves in an adiabatic way in the absence of the energy injection, 
\begin{equation}
E_\mathrm{int}\propto V_\mathrm{b}^{1-\gamma}\propto t^{3\lambda(1-\gamma)}.
\label{eq:Eint1}
\end{equation}
From the assumed self-similarity, the pressure $p_\mathrm{b}$ of the hot bubble is proportional to the post-shock pressure $\propto t^{(2-m)\lambda+m-3}$. 
In the following, we assume that the pressure in the hot bubble is uniform and identical with the post-shock pressure. 
In other words, the entire region in the hot bubble is causally connected, which is justified by the numerical result in Section \ref{sec:results}. 
Therefore, the internal energy of the hot bubble is simply given by the product of the volume of the hot bubble and the internal energy density $p_\mathrm{b}/(\gamma-1)$,
\begin{equation}
p_\mathrm{b}=(\gamma-1)\frac{E_\mathrm{int}}{V_\mathrm{b}}\propto V_\mathrm{b}^{-\gamma}\propto R_\mathrm{fs}^{-3\gamma}\propto t^{-3\lambda\gamma}.
\label{eq:p_b}
\end{equation}
By comparing Equations (\ref{eq:p_f}) and (\ref{eq:p_b}), the exponent $\lambda$ should satisfy the following relation,
\begin{equation}
(2-m)\lambda+m-5=-3\lambda\gamma,
\end{equation}
or equivalently,
\begin{equation}
\lambda=\frac{m-5}{m-3\gamma-2}.
\end{equation}
The requirement $\lambda>1$ leads to $m>3\gamma+2$. 
In other words, the density slope should be sufficiently steep, $m>6$ for $\gamma=4/3$.

We apply this scaling relation to the late-time evolution of model E51, where the hot bubble is expanding into the spherical supernova ejecta with a steep density profile $m=10$ and $\gamma=4/3$. 
In this case, the self-similar expansion law derived above can apply and the exponent $\lambda$ is found to be $\lambda=5/4$. 
Accordingly, the internal energy decays as $E_\mathrm{int}\propto t^{3\lambda(1-\gamma)}=t^{-5/4}$. 

Another case worth considering is the hot bubble expansion in a shallow density slope, $m<3\gamma+2$, as in the inner ejecta, $-d\ln \rho/d\ln r=1<3\gamma+2$, considered in this study. 
In this case, the exponent $\lambda$ is below unity and apparently violates the assumption of the shock propagating outward with respect to the freely expanding medium. 
Namely, the shock front is receding into deeper layers of the ejecta, which is unphysical. 
In fact, under this condition, the pressure $p_\mathrm{b}$ of the hot bubble (equation \ref{eq:p_b}), declines faster than the post-shock pressure $p_\mathrm{f}$ required for the outward shock propagation (equation \ref{eq:p_f}). 
Therefore, the pressure balance will never be reached. 
As a result, the hot bubble is confined in deeper region and passively expands according to the expansion of the ambient ejecta, $\lambda\sim 1$. 

We also note that the assumption of the adiabatic expansion is no longer valid when the radiative diffusion timescale is much shorter than the dynamical timescale of the hot bubble expansion. 
Significant radiative loss at the shock front effectively reduces the adiabatic exponent of the post-shock gas down to unity, $\gamma\rightarrow 1$. 
Accrodingly, the exponent $\lambda$ also approaches unity, $\lambda\rightarrow1$. 
In other words, the hot bubble can also be stuck in the expanding ejecta when the energy loss is considerable.

\end{document}